\documentclass{aa}
\usepackage{epsfig}
\usepackage{amssymb}
 \def\mso{\,\mathrm{M}_\odot}                                                  
 \def\rso{\,{\rm R}_\odot}                                                      
 \def\lso{\,{\rm L}_\odot}                                                      

 \def\kms{\, {\rm km}\, {\rm s}^{-1}}

 \def\h1{\hangindent=1.0truecm \hangafter=0}                                    
 \def\simle{\lesssim}
 \def\simgr{\gtrsim}

 \def\hb{\hfill\break}
 \def\h2{\hb\noindent \hangindent=0.5cm \hangafter=1}

 \def\utw{\smash{\rlap{\lower5pt\hbox{$\sim$}}}}
 \def\udtw{\smash{\rlap{\lower6pt\hbox{$\approx$}}}}

 \def\can{$^{13}$C($\alpha$,n)$\,$}
 \def\nan{$^{22}$Ne($\alpha$,n)$\,$}

\begin{document}
\thesaurus{ 06 (08.05.3; 08.01.1; 08.16.4: 08.18.1)}

\title{Mixing and nucleosynthesis in rotating TP-AGB stars}

\author{N. Langer\inst{1}, A. Heger\inst{2}, S. Wellstein\inst{1} 
  \and F. Herwig\inst{1} }

\institute{
Institut f\"ur Physik, Universit\"at
Potsdam, Am Neuen Palais 10, D--14415~Potsdam, Germany
\and
Astronomy and Astrophysics Department,
    University of California, Santa Cruz, CA~95064
}

\offprints { N. Langer (email: {\tt ntl@astro .physik.uni-potsdam.de})}

\date{Received  ; accepted , }

\maketitle

\begin{abstract}

We present the first evolutionary models of intermediate mass stars 
up to their thermal pulses which include effects of rotation
on the stellar structure as well as rotationally induced mixing of
chemical species and angular momentum. We find a significant angular
momentum transport from the core to the hydrogen-rich envelope and obtain
a white dwarf rotation rate comparable to current
observational upper limits of $\simle 50\,\kms$. 

Large angular momentum gradients at the bottom of the convective envelope
and the tip of the pulse driven convective shell are shown to
produce marked chemical mixing between the proton-rich and the
$^{12}$C-rich layers during the so called third dredge-up.
This leads to a subsequent production of $^{13}$C which is followed
by neutron production through $^{13}$C$(\alpha,$ n$)$ in radiative
layers in between thermal pulses. Although uncertainties in the
efficiency of rotational mixing processes persist, we conclude that
rotation is capable of producing a
$^{13}$C-rich layer as required for the occurrence of the
s-process in TP-AGB stars.

\keywords{Stars: evolution -- Stars: abundances -- Stars: AGB, post-AGB --
Stars: rotation}

\end{abstract}

\section{Introduction}

It is known since long that thermally pulsing Asymptotic Giant Branch
(TP-AGB) stars provide a site for the so called s-process, i.e., the slow 
neutron capture process which forms neutron-rich isotopes heavier than
iron (Clayton 1968). Heavy elements primarily produced by the s-process
are overabundant at the surface of AGB stars (Smith \& Lambert 1990),
including technetium (Little et al. 1987) which has no stable isotope 
and which is produced as $^{99}$Tc ($\tau_{\rm 1/2} = 2.1\times 10^5\,$yr)
in the s-process. In particular the roughly solar magnesium isotopic
pattern found in s-process enriched AGB stars has demonstrated 
that the \can rather than the \nan neutron source is likely to
operate the s-process in AGB stars (Gu\'elin et al. 1995, Lambert et al. 
1995). 

Evidence for {\it in situ} s-processing is found exclusively in carbon stars
(Smith \& Lambert 1990), which correspond to a late evolutionary stage on the
TP-AGB where the stars
have large $^{12}$C enrichments in their envelopes (Iben \& Renzini 1983,
Wallerstein \& Knapp 1998). The $^{12}$C enrichment implies 
that these stars contain,
at certain times, a region at the bottom of their hydrogen-rich envelope
where $^{12}$C is abundant. This region where protons and $^{12}$C
coexist may then perhaps form $^{13}$C through
$^{12}$C(p,$\gamma$)$^{13}$N($\beta^+ \nu$)$^{13}$C.
Although this scenario is unrivaled, the formation of a layer
which is rich in protons {\em and} $^{12}$C in TP-AGB models has 
proven to be difficult, and its existence had hitherto to be assumed
{\it ad hoc} in all s-process calculations (Gallino et al. 1998).
Iben \& Renzini (1982) found a $^{13}$C layer in low metalicity ABG models. 
Recently, Herwig et al. (1997) have obtained a $^{13}$C-rich
layer in TP-AGB models of solar metallicity, 
by invoking a diffusive overshoot layer at convective boundaries.
Here, we investigate for the first time effects of rotationally induced
mixing processes on the TP-AGB.

\section{Numerical method and physical assumptions}

Our calculations have been performed with a hydrodynamic stellar evolution
code (cf., Langer 1998, and references therein), which has been
upgraded to include angular momentum, 
the effect of the centrifugal force on the stellar
structure, and rotationally induced
transport of angular momentum and chemical species due to Eddington-Sweet
circulations, the Solberg-H{\o}iland and Goldreich-Schubert-Fricke instability,
and the dynamical and secular shear instability. We apply the rotational
physics exactly as in Heger et al. (1999). In particular, the effects of
gradients of the mean molecular weight~$\mu$, which pose barriers to any
mixing process, have been included as in Heger et al. (i.e., $f_{\mu}=0.05$).
As in Heger et al., we have also included the effects of $\mu$-barriers
on convection by using the Ledoux-criterion for convection and semiconvection
according to Langer et al. (1983), which is consistent with our treatment
of the rotational mixing (Maeder 1997).

Changes of the chemical composition and the nuclear energy generation rate
are computed using nuclear networks for the three pp-chains, the four
CNO-cycles, and the NeNa- and the MgAl-cycle. Further, the 3$\alpha$-reaction
is included, and ($\alpha$,$\gamma$)-reactions on 
$^{12}$C, $^{14,15}$N, $^{16,18}$O, $^{19}$F, $^{20,21,22}$Ne,
$^{24,25,26}$Mg,  and ($\alpha$,n)-reactions on
$^{13}$C, $^{17}$O, $^{21,22}$Ne, $^{25,26}$Mg.
The inclusion of (n,$\gamma$)-reactions on $^{12}$C, $^{20,21}$Ne,
$^{24,25}$Mg, $^{28,29}$Si allows an order of magnitude estimate
of the neutron concentration. For more details see Heger et al. (1999)
and Heger (1998).

\section{Results}

\subsection{Evolution towards the TP-AGB}

\begin{figure}[t]
\begin{centering}
\epsfxsize=0.9\hsize
\epsffile{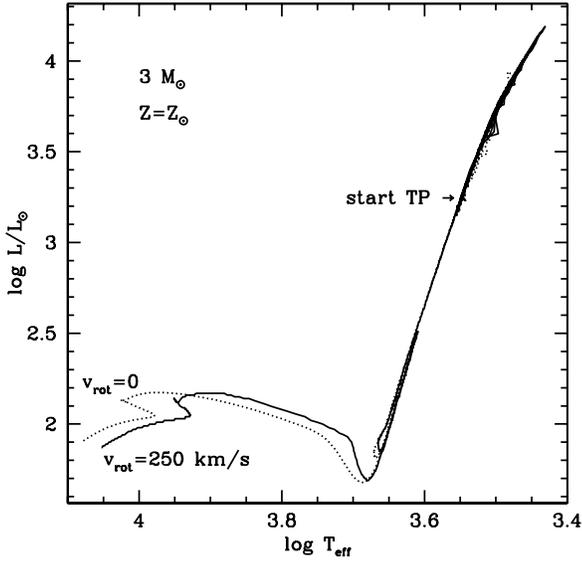}
\caption{Evolutionary track of our rotating 3$\mso$ model (solid line)  
and of a non-rotating reference model (dotted line) in the HR diagram.
The tracks start at the zero age main sequence and end on the TP-AGB.
The beginning of the thermal pulses is marked. 
}
\end{centering}
\end{figure}

\begin{figure}[ht]
\begin{centering}
\epsfxsize=0.7\hsize
\epsffile{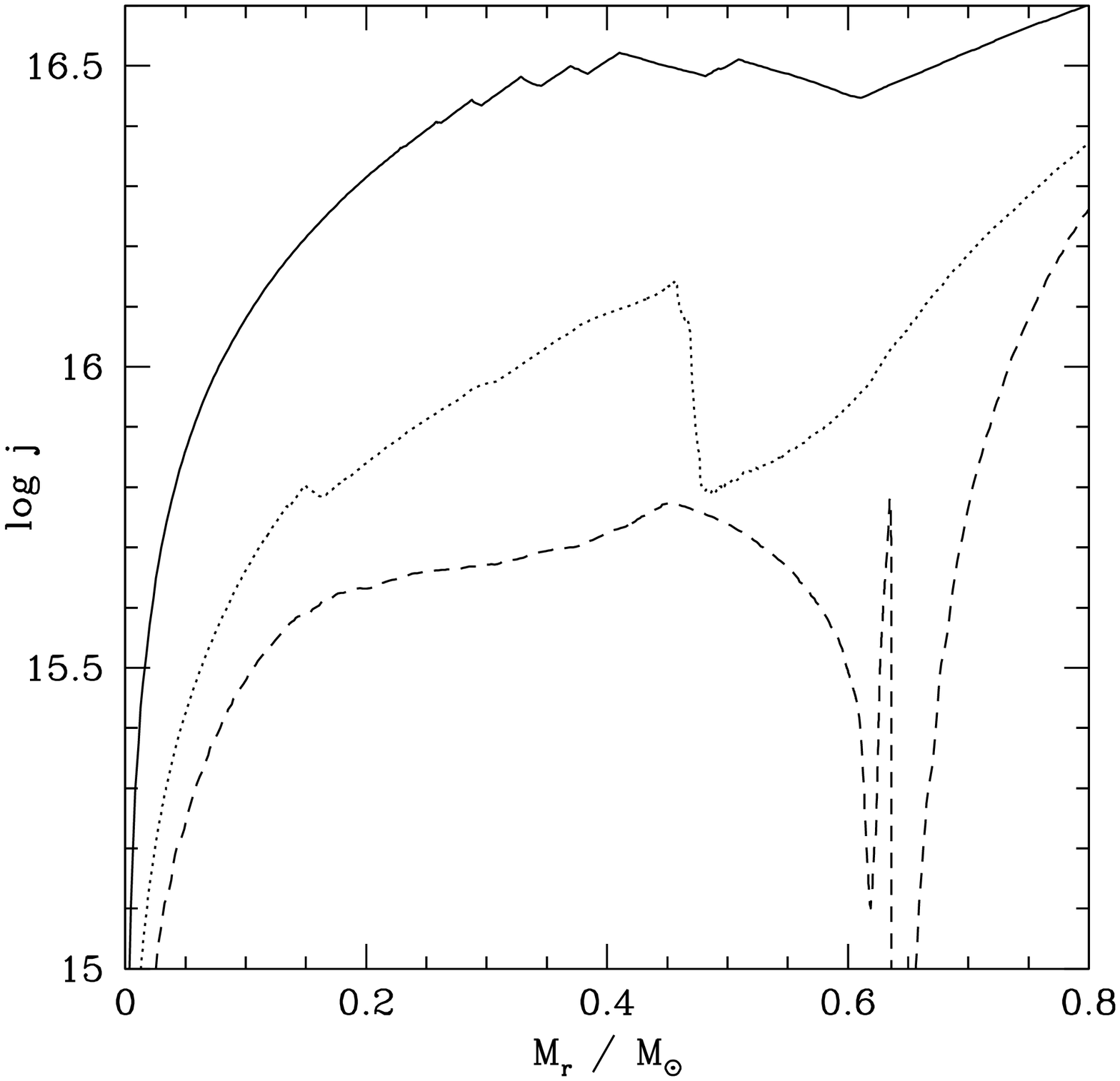}
\epsfxsize=0.7\hsize
\epsffile{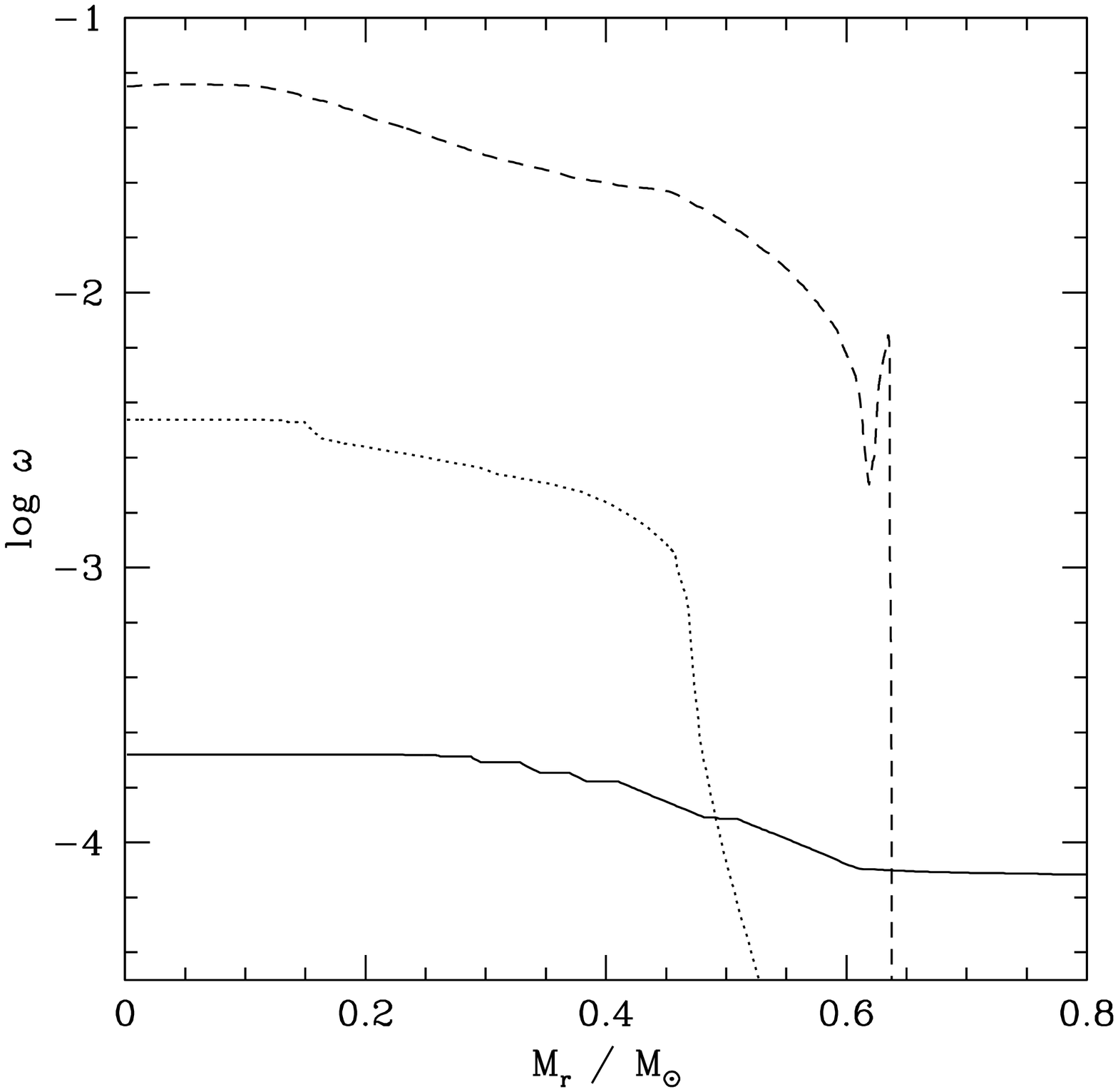}
\caption{{\bf Upper panel:} logarithm of the local specific angular
momentum (in cm$^2\,$s$^{-1}$) as function of the mass coordinate for 
three models of our rotating 3$\mso$ sequence; at core hydrogen
exhaustion ($t\simeq 3.3\times 10^8\,$yr, solid line),
during core helium burning ($t\simeq 3.8\times 10^8\,$yr, dotted line),
and between the 14th and the 15th thermal pulse
($t\simeq 4.4\times 10^8\,$yr, dashed line).
{\bf Lower panel:} logarithm of the local angular velocity (in rad/s)
as function of the mass coordinate for the same models which are displayed
in the upper panel.
}
\end{centering}
\end{figure}

For our pilot study, we chose to compute the evolution of a 3$\mso$
star of roughly solar composition with an initial equatorial rotation
velocity of 250$\kms$, which is typical for late~B main sequence stars
(Fukuda 1982). This choice renders effects of magnetic braking and 
the core helium flash unimportant.
The evolution of our model in the HR diagram, together with that of
a non-rotating reference model, is shown in Fig.~1. 

As in massive stars (Heger et al. 1999), the dominant rotational mixing
process on the main sequence is the Eddington-Sweet circulation.
It leads to a $^{12}$C/$^{13}$C-ratio after the first
dredge-up of~9.4, compared to a value of 19.4 in our non-rotating
model (cf., Boothroyd \& Sackmann 1999). 

Figure~2 sketches the evolution of the angular momentum distribution
in the innermost 0.8$\mso$ of our rotating 3$\mso$ model. It shows that
the core specific angular momentum decreases continuously during the 
evolution. We can give a first quantitative prediction of a white
dwarf rotation rate: $\log (j / {\rm cm}^2 {\rm s}^{-1} ) =15.3$ and 
$R_{\rm WD}=0.01\rso$ yields $v_{\rm rot} = 28\kms$. 
This is of the same order as current observational upper limits
(Heber et al. 1997, Koester et al. 1998).

\subsection{Mixing and nucleosynthesis on the TP-AGB}

\begin{figure}[ht]
\begin{centering}
\epsfxsize=0.9\hsize
\epsffile{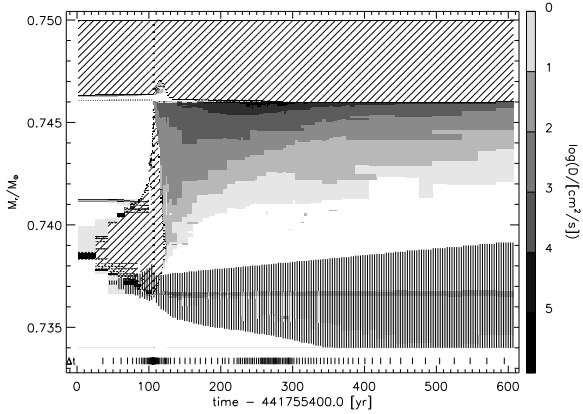}
\caption{Section of the internal structure during and after the 25th 
thermal pulse of our rotating 3$\mso$ sequence. Diagonal hatching denotes
convection. The convective envelope extends down to
$M_{\rm r}\simeq 0.746\mso$. The pulse driven convective
shell is located at $0.737\mso \simle M_{\rm r}\simle 0.746\mso$ and
$30\,{\rm yr}\simle t \simle 120\,$yr.
Vertical hatching denotes regions of significant nuclear
energy generation, i.e., the hydrogen burning shell (at 
$M_{\rm r}\simeq 0.746$ and $t\simle 100\,$yr) and the helium burning
shell ($0.734\mso \simle M_{\rm r}\simle 0.739\mso$ and $t\simgr 40\,$yr). 
Gray shading marks regions of significant rotationally induced 
mixing (see scale on the right side of the figure). Vertical marks at the
bottom of the figure denote the time resolution of the calculation,
where every fifth time step is indicated. Cf. also Fig.~4.
During this thermal pulse, the maximum energy generation rate of the
helium burning shell was $4\times 10^7\lso$. 
}
\end{centering}
\end{figure}

\begin{figure}[ht]
\begin{centering}
\epsfxsize=0.9\hsize
\epsffile{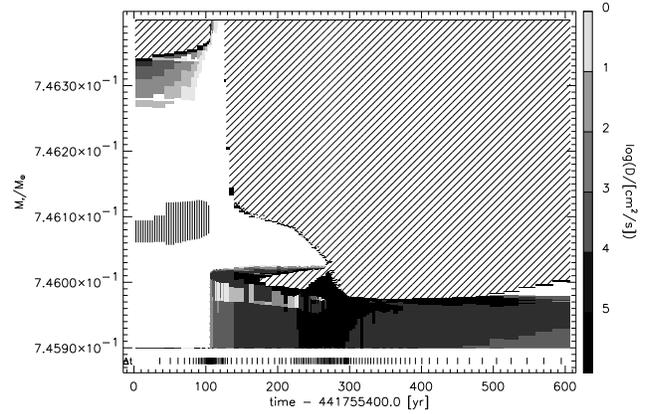}
\caption{Same as Figure~3, for the same time interval, but magnifying
the dredge-up of the convective envelope. Note the hydrogen burning
shell source at $M_{\rm r}\simeq 0.7461$ and $t\simle 100\,$yr, and
the extension of the pulse driven convection zone up to
$M_{\rm r}\simeq 0.74602\mso$ at $t\simeq 100\,$yr.
}
\end{centering}
\end{figure}

Figure~3 shows the evolution of the internal structure during and after
the 25th thermal pulse of our rotating model. It shows that the tip
of the pulse-driven convection zone leaves after its decay a region of 
strong rotational mixing. This mixing becomes even stronger when the
convective envelope extends downward during the third dredge-up event 
(cf. also Figure~4). The reason is that convection enforces close-to-rigid
rotation (cf. Heger et al. 1999), with an envelope rotation rate which is
many orders of magnitude smaller than that of the core. The resulting
strongly differential rotation (cf. also Figure~2) allows the 
Goldreich-Schubert-Fricke instability, and to a lesser extent the
shear instability and Eddington-Sweet circulations, to produce a
considerable amount of mixing between the carbon-rich layer and
the hydrogen envelope.

\begin{figure}[ht]
\begin{centering}
\epsfxsize=0.9\hsize
\epsffile{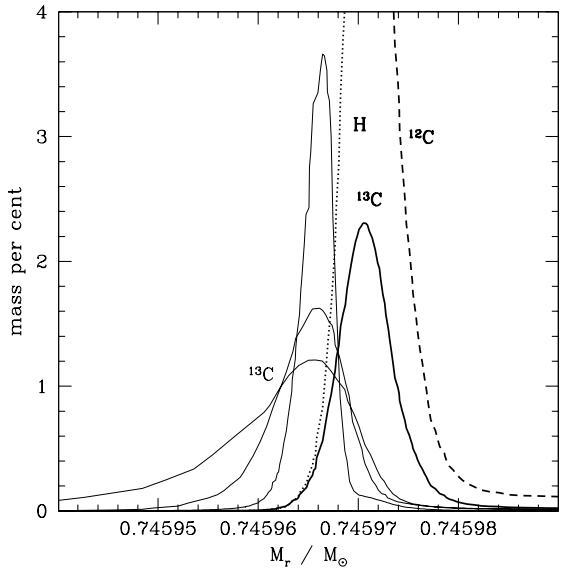}
\caption{Chemical profiles at the location of the maximum depth
of the convective envelope during the 25th thermal pulse 
(cf. Figure~4) of our rotating 3$\mso$ sequence. The dotted and
dashed lines mark the hydrogen and the $^{12}$C mass fractions
at $t=1704\,$yr, with $t=0$ defined as in Figures~3 and~4.
The fat solid line denotes the $^{13}$C mass fraction at the same time.
The three thin solid lines represent the $^{13}$C mass fractions
at $t=2016\,$yr, $t=4155\,$yr, and $t=5139\,$yr, with a later time
corresponding to a smaller peak abundance. The maximum $^{13}$C mass fractions
of 3.6\% occurs at $t=2016\,$yr. The $^{13}$C peak moved inwards
in the time interval from $t=1704\,$yr to $t=2016\,$yr 
due to continued proton captures on both, $^{12}$C and $^{13}$C.
}
\end{centering}
\end{figure}

\begin{figure}[ht]
\begin{centering}
\epsfxsize=0.9\hsize
\epsffile{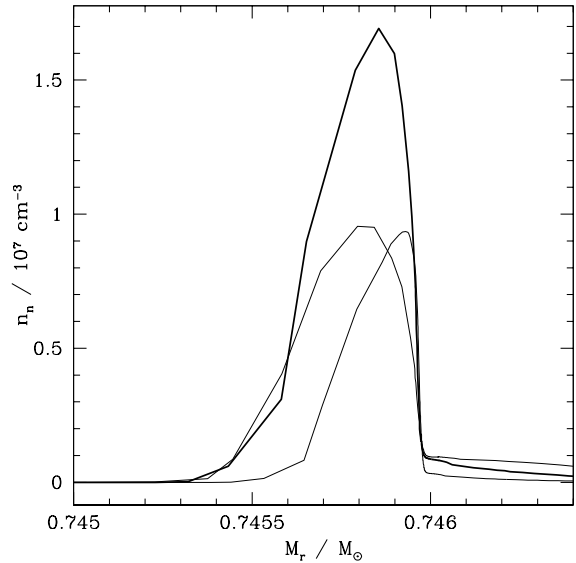}
\caption{Neutron density as function of the mass coordinate for three
models of our rotating $3\mso$ sequence after the 25th pulse. The fat solid
line corresponds to $t=14\, 150\,$yr, the other two (thin solid lines) to
$t=11\, 457\,$yr (peak at largest mass coordinate) and $t=16\, 727\,$yr,
with $t=0$ defined as in Figures~3,~4 and~5. The time span between the
25th and the 26th thermal pulse is $\sim 30\, 000\,$yr.
}
\end{centering}
\end{figure}

Figure~5 depicts the resulting hydrogen and $^{12}$C abundance profiles
after the convective envelope has receded. It shows a layer of
several $10^{-5}\mso$ containing a large mass fraction of protons
and $^{12}$C at the same time. Several 1000$\,$yr after the pulse,
this layer heats up and $^{13}$C is formed through proton capture
on $^{12}$C. Figure~5 shows the resulting $^{13}$C profiles for
four different times. A maximum $^{13}$C mass fraction of almost~4\%
is achieved.

Starting some $10^4\,$yr after the pulse, the $^{13}$C-rich layer
becomes hot enough for $\alpha$-captures on $^{13}$C to occur
(Straniero et al. 1995). Figure~6 shows the resulting neutron
densities $n_{\rm n}$ for three different times. 
Note that we did not include the
reaction $^{14}$N(n,p)$^{14}$C in our network. Although most of the
resulting protons may form new $^{13}$C, it may be an effective
neutron sink (Jorissen \& Arnould 1989), in particular as also a large
abundance of (primary) $^{14}$N is produced in the $^{13}$C-rich layer.
Thus, our neutron densities can only be considered as an order of
magnitude estimate. With $n_{\rm n}\simeq 10^7\,$cm$^{-3}$ for
$\sim 10^4\,$yr, we obtain a neutron irradiation of 
$\tau\simeq 10^{27}\,$neutrons/cm$^2$ 
which results roughly in a number of neutron
captures per iron seed of $n_{\rm c}\simeq 75$, i.e. a main component
s-process (cf. Figures~7.22 and 7.23 of Clayton, 1968).  

\section{Discussion}

By applying the concept of rotationally induced mixing as it has
been developed for massive stars in our group during the last years
without alteration to a $3\mso$ TP-AGB model sequence, we obtain conditions
which appear favorable for the development of the s-process,
i.e. a $^{13}$C-rich layer which produces a considerable neutron flux 
later-on. Although our model develops only a very late and weak third
dredge-up we believe that the mechanism which diffuses the protons
into the carbon layer and $^{12}$C into the envelope must occur 
with a similar magnitude in all
TP-AGB stars which develop a third dredge-up. The reason is that the huge
specific angular momentum jump at the hydrogen/carbon interface
--- five orders of magnitude in our case --- is independent of the
depth of the third dredge-up.

The maximum $^{13}$C
abundance and its distribution in our model is, at first, 
similar to that found due to diffusive convective overshooting
by Herwig et al. (1997). However, in our case
the rotational mixing spreads the $^{13}$C peak out before the neutrons
are produced (cf. Figures~5 and~6), which is not the case
in the models of Herwig et al. (1997). At the present time we can not
discriminate which of these scenarios would agree better with 
empirical constraints. However, we want to stress that both
mechanisms of $^{13}$C production, rotation and overshooting, 
do not exclude each other, and that it is possible that they
act simultaneously in AGB stars.

Finally, we want to emphasize that,
although stars of less than $\sim 1.3\mso$ 
lose 99\% of their angular momentum due to a magnetic wind during
their main sequence evolution, 
it can not be excluded that the proposed mechanism of $^{13}$C-production
due to differential rotation also works for them.
Certainly, the sun's core will spin-up and the envelope will further spin down
during its post-main sequence evolution, which may result in
a specific angular momentum jump of similar magnitude.
The investigation of the mass and metallicity dependence of the 
production of $^{13}$C due to rotation is an exciting task for the near future.

\begin{acknowledgements}
  We are grateful to Thomas Bl\"ocker for many fruitful discussions.
  This work has been supported by the Deutsche Forschungsgemeinschaft
  through grants La~587/15-1 and 16-1.
\end{acknowledgements}

\end{document}